\documentclass{article}

\usepackage[final]{neurips_2025}

\usepackage[utf8]{inputenc} 
\usepackage[T1]{fontenc}    
\usepackage{hyperref}       
\usepackage{url}            
\usepackage{booktabs}       
\usepackage{amsfonts}       
\usepackage{nicefrac}       
\usepackage{microtype}      
\usepackage{xcolor}         

\usepackage{times}
\usepackage{soul}
\usepackage{url}
\usepackage{hyperref}

\usepackage[small]{caption}
\usepackage{graphicx}
\usepackage{amsmath}
\usepackage{amsthm}
\usepackage{booktabs}
\usepackage{algorithm}
\usepackage{algorithmic}
\usepackage{lineno}

\usepackage{algorithm}
\usepackage{algorithmic}
\usepackage{subcaption}
\usepackage{tcolorbox}
\usepackage{pifont}
\usepackage{multirow}
\usepackage{rotating} 
\usepackage{array}
\usepackage{graphicx}
\usepackage{xcolor}
\usepackage{soul}
\usepackage{amsmath}
\usepackage{booktabs}
\usepackage{tikz}
\usepackage{amsfonts}
\usepackage{wrapfig}

\usepackage{colortbl}

\sethlcolor{yellow}

\definecolor{yellow}{rgb}{1, 1, 0.5}
\definecolor{green}{rgb}{0.75, 1, 0.75}
\definecolor{lightblue}{rgb}{0.68, 0.85, 0.9}

\usepackage{newfloat}
\usepackage{listings}

\newcommand{\highlight}[1]{{\noindent\textbf{#1}}}

\title{CoreGuard: Safeguarding Foundational Capabilities of LLMs Against Model Stealing in Edge Deployment}

\author{
\textbf{Qinfeng Li}$^{1}$\hspace{0.4em}
\textbf{Tianyue Luo}$^{1,2}$\hspace{0.4em}
\textbf{Xuhong Zhang}$^{1,3}$\thanks{~Corresponding to \texttt{zhangxuhong@zju.edu.cn}} \hspace{0.4em}
\textbf{Yangfan Xie}$^{1}$\hspace{0.4em}
\textbf{Zhiqiang Shen}$^{1}$\\[2pt]
\textbf{Lijun Zhang}$^{4}$\hspace{0.4em}
\textbf{Yier Jin}$^{5}$\hspace{0.4em}
\textbf{Hao Peng}$^{6}$\hspace{0.4em}
\textbf{Xinkui Zhao}$^{1}$\hspace{0.4em}
\textbf{Xianwei Zhu}$^{7}$\hspace{0.4em}
\textbf{Jianwei Yin}$^{1}$\\[4pt]
$^{1}$School of Software Technology, Zhejiang University \\
$^{2}$Institute of Software, Chinese Academy of Sciences \\
$^{3}$Ningbo Global Innovation Center, Zhejiang University\\
$^{4}$Washington University in St. Louis \\
$^{5}$University of Science and Technology of China \\
$^{6}$College of Computer Science and Technology, Zhejiang Normal University \\
$^{7}$China Electronics Technology Design and Research Institute \\
}

\begin{document}

\maketitle

\begin{abstract}
Proprietary large language models (LLMs) exhibit strong generalization capabilities across diverse tasks and are increasingly deployed on edge devices for efficiency and privacy reasons. However, deploying proprietary LLMs at the edge without adequate protection introduces critical security threats. Attackers can extract model weights and architectures, enabling unauthorized copying and misuse. Even when protective measures prevent full extraction of model weights, attackers may still perform advanced attacks, such as fine-tuning, to further exploit the model. Existing defenses against these threats typically incur significant computational and communication overhead, making them impractical for edge deployment.
To safeguard the edge-deployed LLMs, we introduce CoreGuard, a computation- and communication-efficient protection method. CoreGuard employs an efficient protection protocol to reduce computational overhead and minimize communication overhead via a propagation protocol. Extensive experiments show that CoreGuard achieves upper-bound security protection with negligible overhead.
\end{abstract}

\section{Introduction}
\label{Introduction}

Large language models (LLMs), especially proprietary ones, such as ChatGPT~\cite{openai2023chatgpt} and Claude~\cite{anthropic2023claude}, demonstrate exceptional generalization ability across various tasks~\cite{mm1.5,mm1.2}. 
Additionally, deploying LLMs on edge devices is a growing trend for latency- and privacy-sensitive tasks, e.g., Apple Inc. introduced Apple Intelligence, which integrates a 3-billion-parameter LLM into users' devices in the latest iOS version~\cite{aaai1.1}.
However, when these proprietary LLMs are deployed to edge devices without adequate protection, adversaries can extract detailed model information (including architecture and weights) through software analysis techniques~\cite{mm1.7,mm1.8}, leading to unauthorized copying and misuse outside the intended device. Even if some protections prevent attackers from fully extracting the original weights, attackers can still perform more advanced attacks, such as fine-tuning the partially recovered model to exploit its embedded knowledge and strong generalization capabilities for new tasks. We refer to this threat as \textit{foundational capability stealing}. This threat is especially practical for proprietary, domain-specific LLMs trained on private data, like BloombergGPT~\cite{wu2023bloomberggpt} in finance or Med-PaLM 2~\cite{singhal2025toward} in healthcare, where comparable open-source alternatives are limited.
Considering the substantial resources required to develop high-performance LLMs~\cite{mm1.28}, it is crucial to ensure robust protection against these threats in edge deployments.

Unfortunately, as shown in Table \ref{table_intro}, traditional solutions struggle to protect the edge-deployed LLMs.
Specifically, passive protection methods, such as watermark~\cite{mm1.18,mm1.19,mm1.20}, are not applicable since only the proof of ownership is insufficient in such an unsupervised edge operation scenario, where attackers can misuse the model without detection.
In contrast, active protection works by allowing only authorized users to use the well-performed model. For example, some work encrypts models before deploying them on devices~\cite{mm1.25,mm1.26}, and these models are only decrypted before execution.
However, it's crucial to recognize that while these solutions can implement effective protection before the inference state, current studies~\cite{mm1.7,mm1.8} suggest that, even after decryption, models remain susceptible to runtime attacks during inference, i.e., attackers reverse engineer the models in their runtime state.

\begin{table}[t]
\caption{Comparison with existing solutions. \checkmark/\ding{55} indicate whether each property is satisfied.
}
\centering
\resizebox{0.85\columnwidth}{!}{
\begin{tabular}{lccccc}
\toprule
Solutions (exemplar)   & Proactivity & Runtime Security & Backbone Protection  & \textbf{Sufficiency} & \textbf{Efficiency}\\ \midrule
Watermarking~\cite{mm1.18}        & \ding{55} & \ding{55}  & \ding{55} & \ding{55} & \checkmark    \\
Model encryption~\cite{mm1.25}         & \checkmark & \ding{55} & \ding{55} & \ding{55} & \checkmark    \\
TPTE~\cite{mm0.0}        & \checkmark & \checkmark  & \ding{55} & \ding{55} & \ding{55}    \\
\textbf{PPTE}~\cite{mm4.12}   & \checkmark & \checkmark & \checkmark     & \ding{55}  & \ding{55}  \\
\textbf{PSP}~\cite{mm4.13}             & \checkmark & \checkmark  & \checkmark & \checkmark & \ding{55}   \\
CoreGuard~(ours)    & \checkmark & \checkmark & \checkmark & \checkmark & \checkmark                 \\ \bottomrule
\end{tabular}
}
\label{table_intro}
\end{table}

To defend against runtime attacks, one potential solution~\cite{mm1.23,mm1.24} is to place the model into a secure execution environment, e.g., a trusted execution environment (TEE), which are typically implemented as a CPU-based enclave (e.g., ARM TrustZone, Intel SGX) that stores sensitive data and safeguards against runtime attacks. However, directly placing the entire model within a TEE is impractical, as it results in approximately a 50$\times$ reduction in model efficiency due to the TEE's limited computational speed~\cite{mm5.11}. 
Thus, some researchers propose only putting the most critical parameters in TEEs and offloading the rest computation to GPUs. For example, Zhang et al.~\cite{mm0.0} protect only task-related adapters within TEE and offload the model backbone to GPUs.
However, such approaches are primarily effective for traditional task-specific models and are insufficient for protecting LLMs, as they leave the model backbone directly exposed.

To protect the backbone, a straightforward idea~\cite{mm4.12,mm4.14} is to place a subset of the backbone (e.g., the last layer) in the TEE, i.e., Partial Parameter TEE Execution (\textbf{PPTE}).
However, prior work~\cite{mm0.0} shows that PPTE only provides insufficient protection, and this limitation becomes even more critical when applied to LLMs.
Specifically, PPTEs crudely execute weights in TEE for protection, where the limited computational power of TEEs restricts the number of protected weights. The scarcity of protected weights makes it easy for attackers to reconstruct them, even with just 1\% of the training datasets, compromising security~\cite{mm0.0}. 
Even worse, if an attacker aims to exploit a LLM’s foundational capabilities, their target task is likely one for which they already have abundant labeled data (e.g., 100\% training set), making theft easier.

To increase the number of protected weights, a promising approach is to protect weights through shuffling, i.e., Parameter Shuffling Protection (\textbf{PSP})~\cite{mm4.13,aaai0.0}.
For example, ShadowNet~\cite{mm4.13} protects model weights by shuffling the channels of convolutional kernels.
The protection ensures that only the corresponding shuffled input can be correctly computed with the shuffled weights.
This input-shuffling process is performed within the TEE, thus ensuring its security.
However, the excessive data transfer overhead between the TEE and GPU makes ShadowNet impractical for LLMs.
Specifically, each shuffled layer requires transferring its input from the GPU to the TEE and back, resulting in 448 TEE-CPU transfers for a LLaMA3-8B model with 224 linear layers (each linear layer requires 2 transfers) to generate a single token. Therefore, with an input of 128 context length, each transfer would average 3MB of data (assuming float-32 precision), leading to a total data transfer volume of approximately 1.3GB (448 $\times$ 3MB).
Given that mainstream mobile platform TEEs, like TrustZone, have a transfer rate of about 1GB/s between TEE and GPU~\cite{mm2.9}, generating a single token takes about 1.3 seconds. Consequently, producing a complete output would require several hundred seconds (assuming it consists of 100 tokens) solely for data TEE-GPU transfer.

In summary, maintaining acceptable computation and communication overhead under sufficient security of LLM in edge deployment is an unresolved challenge for existing solutions. To address this, we propose CoreGuard, a computation- and communication-efficient approach designed to prevent the model from working without the proper authorization from the trusted hardware, i.e., TEE, within the edge device.
To reduce TEE execution overhead, CoreGuard is inspired by prior PSP solutions by securing parameters through obfuscation, which allows model computations to be performed on the GPU.
Specifically, it employs a \textit{protection protocol} that row-permutes the weight matrices of linear layers, ensuring their input features must be correspondingly column-permuted (i.e., authorization) by TEE.
Crucially, to avoid requiring TEE authorization for each linear layer and minimize TEE-GPU transfer overhead, CoreGuard introduces a \textit{propagation protocol} that reduces TEE authorizations to a single initial authorization. After this, all subsequent protected layers apply column permutations to their outputs, enabling the initial authorization to be propagated.

Our evaluation shows that CoreGuard outperforms existing defenses in security and efficiency. 
Besides, the experimental results show no difference in accuracy between the CoreGuard-protected model and the original model. The contributions of this work are as follows:
\begin{itemize}
\item We are the first to address the protection of foundational capabilities in edge-deployed LLMs. Our work systematically characterizes the security challenges in this setting and identifies the requirements for the protection of edge-deployed LLMs.
\item We propose CoreGuard, a plug-and-play solution that utilizes a lightweight authorization mechanism to protect edge-deployed LLMs. It employs a propagation protocol, significantly reducing transfer overhead while maintaining a low computation overhead.
\item Extensive experiments demonstrate that compared to the existing solutions, CoreGuard offers a higher security guarantee with lower overhead and no accuracy loss. 
\end{itemize}

\section{Threat Model}
In this paper, we consider two parties: the defender and the attacker. The defender is the party that owns the edge-deployed model. The attacker aims to steal the model.

\highlight{Defender's Goal.} 
The defender aims to deploy a locked model on the device, ensuring it works only with proper authorization from the trusted hardware (i.e., TEE) within the device. The defender can control its model and modify it to ensure protection. This protection ensures that, when correctly authorized, the model permits normal queries from authorized users, while other attacks based on these legitimate queries (e.g., distillation attacks) are orthogonal to our work.

\highlight{Adversary's Goal.} 
The attacker aims to abuse the deployed model off-device (i.e., without TEE authorization) for their task. A straightforward way is to try to reverse the authorization process so that the locked model can be used independently of the device. Another more practical new way is to fine-tune the locked model to obtain a model that excels at a desired task~\cite{aaai0.0,mm0.0}.

\highlight{Adversary's Capability.} 
The attacker could \textbf{\textit{decide the target task}} and \textbf{\textit{possess sufficient well-labeled data}}, whereas prior work often assumes access to only 1\% of the dataset~\cite{mm0.0}.
In this paper, we consider TEE to be a secure world, while other hardware (e.g., GPU and CPU) could be white-box exposed to attackers.
Therefore, the attacker can have access to the details, e.g., model architecture and weights, of the locked model outside TEE.

\section{Design of CoreGuard}
\label{Design of CoreGuard}
This section presents our proposed protection method, CoreGuard, which utilizes a permutation strategy to address the key requirements outlined in Section \ref{Introduction}.


\subsection{Approach Overview}
\label{Approach Overview}
\begin{figure*}[t]
\centering
\includegraphics[width=1\textwidth]
{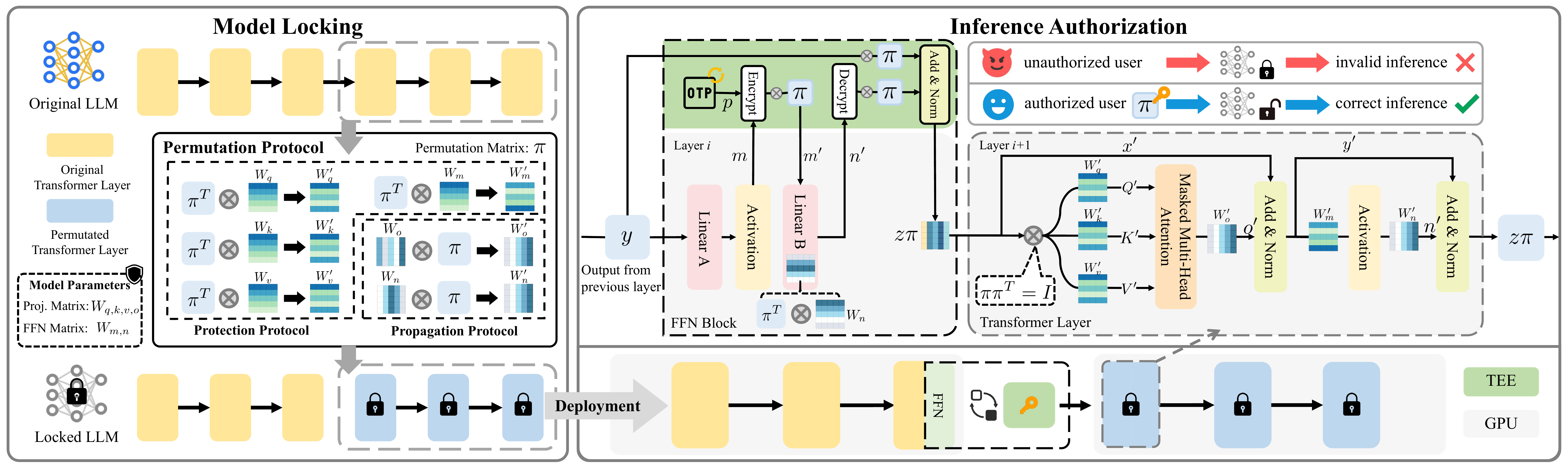} 
\vspace{-4.5mm}
\caption{An overview of CoreGuard. (a) Model locking: before deployment, CoreGuard permutes layers in the original model, thus creating a locked model. 
(b) Inference authorization: during inference, the input feature of the permuted layers is authorized, which is integrated within the FFN block of the preceding transformer layer.}
\vspace{-3mm}
\label{figure_method}
\vspace{-0.8\baselineskip}
\end{figure*}

As shown in Figure \ref{figure_method}, CoreGuard operates in two phases: model locking (before deployment) and inference authorization (post-deployment).
In the \textit{model locking phase}, CoreGuard locks a trained model by applying a \textbf{protection protocol} to the weights of linear layers, i.e., swapping rows of the weight matrix.
These row permutations act as a \textit{lock}, rendering the linear layers dysfunctional, thus making the overall model unusable. These locked layers can only function properly with inputs that are correspondingly column-permuted, which essentially acts as \textit{authorization}. However, directly using a TEE to authorize each locked layer would result in significant TEE-GPU transfer overhead. To address this, CoreGuard proposes a \textbf{propagation protocol}, which enables the features to be column-permuted by the network itself. Specifically, CoreGuard permutes the columns of certain layers, thus through these layers' operation, their output features are column-permuted, which achieve authorization similarly. In this way, the TEE only needs to manage the initial authorization, and the authorization can be propagated to all subsequent layers.

The \textit{inference authorization phase}, as shown in the black dashed box of Figure~\ref{figure_method} (b), aims to securely perform initial authorization.
A naive method directly authorizes the original output $z$ in the TEE and returns $z\pi$, but this obviously will leak $\pi$ via input-output comparisons. 
One solution is to place both Linear B and the add-norm layer inside the TEE, fully hiding $z$, but executing Linear B in the TEE introduces high overhead.
To reduce cost, we offload linear B to the GPU.
However, exposing Linear B's output $n$ will leak $z$.
Therefore, we apply OTP noise $p$ to Linear B’s input $m$.
Since OTP also requires protection, we apply a permutation after encryption, producing $m'$ (i.e., $(m + p)\pi$). Correspondingly, Linear B’s weights are pre-permuted by $\pi^T$ to offset $\pi$. Overall, as shown in the Figure \ref{figure_method} (b), the feature enters the TEE twice: first for OTP encryption (encrypting $m$ to $m'$), and second for performing authorization to produce $z\pi$.

\newcommand{\graycomment}[1]{\textcolor{gray}{// #1}}
\newcommand{\gray}[1]{ \textcolor{gray}{// #1}}
\subsection{Model Locking}
Given a classic transformer model, we describe how to lock a transformer layer within it. We first apply the protection protocol to layers involved in input feature projection (e.g., the QKV projection layer) to secure the model. Then, the propagation protocol is applied to layers managing output projection (e.g., the output projection layer in the attention block and the FFN block). Finally, we demonstrate the functionality of the locked transformer layer.

\highlight{Transformer Layer Formalization.}
We begin by formalizing a standard transformer layer. Let $x$, and $z \in \mathbb{R}^{l \times d}$ denote its input and output, where $l$ is the sequence length (e.g., the number of tokens) and $d$ is the model dimension. We define a classic transformer layer as a function $f_{w}: \mathbb{R}^{l \times d} \to \mathbb{R}^{l \times d}$ with weight parameters $w$. The transformer layer, i.e., $f_{w}(x) = z$, is computed as follows:
\begin{equation}
\begin{aligned}
Q &= xW_q, K = xW_k, V = xW_v,&\graycomment{\textit{QKV project}}\\
o &= \text{softmax}\left(\frac{QK^T}{\sqrt{d/h}} + M\right)VW_o, & \graycomment{\textit{Attention}}\\
y &= \gamma_1 \odot \frac{o + x - \mu_{o+x}}{\sigma_{o+x}} + \beta_1,   &  \graycomment{\textit{Add \& norm}}\\
m &= \text{ReLU}(yW_m + b_m),    &  \graycomment{\textit{FFN input}}\\
n &= mW_n + b_n    &  \graycomment{\textit{FFN output}}\\
z &= \gamma_2 \odot \frac{y + n - \mu_{y+n}}{\sigma_{y+n}} + \beta_2,    & \graycomment{\textit{Add \& norm}}\\
\end{aligned}
\end{equation}
where $w$ includes the attention weights $W_q$, $W_k$, $W_v$, and $W_o$ $ \in \mathbb{R}^{d \times d}$, the add-norm weights $\gamma_1$, $\beta_1$, $\gamma_2$, and $\beta_2$ $ \in \mathbb{R}^d$, the FFN weights $W_m$ and $W_n$ $ \in \mathbb{R}^{d \times d}$, the bias $b_m$ and $b_n$ $ \in \mathbb{R}^d$. $h$ is the number of attention heads. The mask $M \in \mathbb{R}^{d \times d}$ is an all-zero matrix in the encoder and has negative infinity in its upper right corner in the decoder. 
Notably, among these layers, we refer to $W_q$, $W_k$, and $W_v$ in the attention block, and $W_m$ in the FFN block as \textit{input-processing layers} as they process the input feature of their blocks. We refer to $W_o$, $W_n$, and the add-norm weights in each block as \textit{output-processing layers} as they manage the output of the blocks.

\highlight{Protection Protocol.}
To achieve protection, we row-permutate the \textit{input-processing layers} for protection. These layers process the module's inputs directly, thus they have the ability to cause computation failures if the input is not authorized, leading to incorrect results.

Specifically, let $\pi \in \{0, 1\}^{d \times d}$ denote a permutation matrix, where $\forall \pi,\pi \pi^T = I$, where $I$ is the identity matrix, a property of the permutation matrix. We row-permute the parameters $w$:
\begin{align}
\begin{split}
W'_q = \pi^T W_q, \quad W'_k = \pi^T W_k, \quad W'_v =  \pi^T W_v, \quad W'_m =  \pi^T W_m.\\
\end{split}
\end{align}
Thus, only the corresponding column-permuted input can be computed with these layers. For instance:
\begin{align}
\begin{split}
x \pi \pi^T W_q = x W_q = Q,
\end{split}
\end{align}
where input's column permutation (i.e., $\pi$, the authorization) and row permutations (i.e., $\pi^T$, the lock) offset each other ($\pi \pi^T = I$), resulting in the same computation as the original.

\highlight{Propagation Protocol.}
The propagation protocol aims to avoid repeated TEE authorization by allowing each transformer layer to automatically receive an authorized input $z\pi$ from the previous layer. Specifically, the output of a transformer layer is directly determined by four output-processing layers: $W_o$, $W_n$, and two add-norm layers. Once all their outputs are column-permuted, the overall output is column-permuted, thus achieving automatic authorization. Therefore, the problem simplifies to column-permuting the output of a single layer. For instance, as illustrated below, column-permuting $W_n$ to $W_n\pi$ transforms its original output $n$ into a column-permuted output $n\pi$:
\begin{align}
\begin{split}
n' = mW'_n + b'_n = mW_n\pi + b_n\pi=n\pi,
\end{split}
\end{align}
Therefore, to implement propagation protocol, we column-permutate all \textit{output-processing layers}, ensuring the feature can be re-authorized before exiting the module:
\begin{align}
\begin{split}
W'_o = W_o \pi, \,\,\,\, \gamma'_1 = \gamma_1 \pi, \,\,\,\, \beta'_1 = \beta_1 \pi, \,\,\,\,
W'_n = W_n \pi, \,\,\,\, b'_n = b_n\pi, \,\,\,\, \gamma'_2 = \gamma_2 \pi,  \,\,\,\, \beta'_2 = \beta_2 \pi.
\end{split}
\end{align}
\highlight{Locked Transformer Layer Formalization.}
With the permuted weights (denoted as $w'$), taking $x \pi$ as authorized input, its functionality can be described as follows:
\begin{align}
\begin{split}
Q' &= x \pi W_q' = x W_q = Q, \\
K' &= x \pi W_k' = x W_k = K, \\
V' &= x \pi W_v' = x W_v = V, \\
o' &= \text{softmax}\left(\frac{Q'K'^T}{\sqrt{d/h}} + M\right)V'W_o' = o \pi, \\
y' &= \gamma_1' \odot \frac{o' + x\pi - \mu_{x\pi+o'}}{\sigma_{x\pi+o'}} + \beta_1' = y \pi. \\
m' &= \text{ReLU}(y' W_m' + b_m) = m. \\
n' &= m'W_n' + b_n' = n \pi,\\
z' &= \gamma_2' \odot \frac{n' + y' - \mu_{y'+n'}}{\sigma_{y'+n'}} + \beta_2' = z \pi. \\
\end{split}
\end{align}
Thus, the functionality of the locked layer can be represented as \( f_{w'}(x') = z\pi = f_{w}(x)\pi \), valid only when \( x' = x\pi \), thereby preventing unauthorized access (without $\pi$). 
When the permuted transformer layer receives authorized input ($x\pi$), its output ($z\pi$) matches the original output ($z$) with a column permutation, authorizing the next layer. This propagation is consistent across all subsequent layers. Therefore, authorization is required only for the first permuted layer.

\subsection{Inference Authorization}
\label{Inference Authorization}
During the inference, the initial column-permutated feature ($x\pi$) is generated while ensuring the security of this authorization process.
As shown in Figure \ref{figure_method}, TEE encrypts the feature before the FFN block's output linear layer. After the GPU processes the feature with this layer, it re-enters the TEE for decryption. Finally, the add-norm computation is performed, followed by the permutation of the output before it is returned.
The detailed descriptions are provided in the following subsections.

\highlight{Encryption.}
At the beginning, the input linear layer of the FFN block receives $y $ as input from the previous layer:
\begin{align}
\begin{aligned}
\label{mlp_first_linear}
m = \text{ReLU}(y W_m + b_m). \\
\end{aligned}
\end{align}
Following this, before the output linear layer, the feature is encrypted using an OTP. Additionally, to protect this encryption process from input-output differencing, we introduce positional obfuscation by applying a permutation:
\begin{align}
\begin{aligned}
\label{mlp_encryption}
m' = m\pi + p \pi,\\
\end{aligned}
\end{align}
where $p$ is the pad, a noise matrix of the same shape as $m$, and $m'$ is the encrypted permuted feature. The principle of the one-time pad (OTP)~\cite{mm2.10} ensures that $p$ is different each time, thus even for the same $m$, $m'$ produced will differ. In this way, the OTP encryption (i.e., $p$) and the permutation (i.e., $\pi$) can conceal each other.

\highlight{Output Linear Layer.}
The FFN block’s output linear layer processes the encrypted feature. However, since $m'$ is permuted, the layer’s parameters must be pre-aligned to ensure correct computations. Specifically, we correspondingly permute the output linear layer ($W_n$) \textit{before deployment}:
\begin{align}
\begin{split}
\label{mlp_permutation}
 W'_n =  \pi^T W_n.
\end{split}
\end{align}
With the above preparation, \textit{during inference}, the encrypted feature $m'$ is transferred to GPUs and processed by the permuted output linear layer $W_n'$:
\begin{align}
\begin{aligned}
\label{mlp_seconed_linear}
n' = m' W'_n = (m\pi + p \pi) \pi^T W_n + b_n = n + p W_n. \\
\end{aligned}
\end{align}

\highlight{Decryption.}
If the OTP noise (i.e., $p W_n$) is not eliminated, the network's functionality is compromised. Specifically, OTP implementation must meet two requirements: 1) conceal both the encryption and decryption; 2) all computations on the feature before decryption are linear. 
To meet these requirements, $n'$ is transferred to the TEE for decryption: 
\begin{align}
\begin{aligned}
\label{mlp_decryption}
&n'' = n' - p W_n = n.\\
\end{aligned}
\end{align}
Notably, following prior work~\cite{mm5.11}, $p W_n$ can be conducted by the model provider or in an offline phase. 
Both strategies do not increase the overhead of online inference or impede its efficiency~\cite{mm0.0}.

\highlight{Authorization.}
Lastly, the decrypted feature is processed by the add-norm layer and permuted to achieve authorization in TEE. The steps are as follows:
\begin{align}
\begin{aligned}
\label{mlp_authorization}
z' &= (\gamma_2 \odot \frac{n + y - \mu_{y+n}}{\sigma_{y+n}} + \beta_2)\pi  = z\pi, \\
\end{aligned}
\end{align}
where $z'$ (i.e., $z \pi$) is the authorized feature, which will be the input feature for the subsequent permuted transformer layer, thus achieving authorized usage. Notably, the TEE \textit{only authorizes once for each inference}, minimizing communication overhead by limiting TEE-GPU transfers to 5 rounds. Furthermore, it uses only lightweight computations (e.g., matrix addition), ensuring minimal computation overhead.

\highlight{Authorization Position.}
The TEE authorization position, which determines the number of layers to be locked, is a hyperparameter.
CoreGuard sets this position to the midpoint to enhance security, as detailed in Appendix \ref{Authorization Position}.
Specifically, when the authorization point is in the middle, attackers must recover more parameters, increasing the difficulty of theft. 
Essentially, permutation only maps parameters to a new domain without disrupting their functionality, meaning the attacker only needs to recover the missing layers, whether in the original or the new domain, to obtain a complete model. If the authorization occurs at the beginning or the end, attackers can retrain just one layer to recover, which is similar to prompt tuning or training a classification head.
In contrast, placing it in the middle requires the attacker to restore at least half of the parameters, which is more difficult.

\highlight{Security Analysis.}
Potential attackers might attempt to steal the locked model by recovering the permuted parameters. However, it is impossible as the probability of guessing the correct $\pi$ is $1/(d!)$. In practice, $d$ is typically larger than 512, e.g., $d = 4096$ in LLaMA3-8B~\cite{mm4.8}.

Alternatively, attackers may attempt to recover $\pi$ by exploiting the TEE's functionality—for instance, by trying to crack $\pi$ from the TEE’s inputs and outputs.
However, accurately solving $\pi$ is infeasible, as the problem is ill-posed. Specifically, even with auxiliary information, the task reduces to solving a Learning With Errors (LWE) problem, which is widely regarded as NP-hard (see Appendix~\ref{NP}). 

Although solving $\pi$ exactly is impossible, attackers might try to approximate TEE's functionality to facilitate model stealing. For example, trying to learn a mapping from $y$ to $z\pi$ to bypass OTP encryption. However, this approach is also ineffective (as shown in Appendix \ref{Adaptive Attack}): the mapping is nonlinear, involves a massive number of parameters, and even minor approximation errors can invalidate the result~\cite{mm4.9}.

\section{Experiments}
In this section, we perform extensive experiments to answer the following research questions (\textbf{RQ}s):
\vspace{-1.5mm}
\begin{tcolorbox}[colback=gray!20!white,colframe=black,left=2mm, right=2mm, top=1mm, bottom=1mm, boxsep=0pt]
\footnotesize
\textbf{RQ1:} How secure is CoreGuard, and does it effectively protect the foundational capabilities of LLMs?
\textbf{RQ2:} How does CoreGuard's computation and communication overhead compare to other defenses?
\textbf{RQ3:} Does CoreGuard sacrifice the accuracy of the model?
\end{tcolorbox}


\subsection{Experimental Settings}
\label{Experimental Settings}
\highlight{Datasets.} To evaluate CoreGuard, we assume the attacker attempts to steal the LLM to different target tasks, including four domain-specific tasks: GSM8k (mathematics)~\cite{mm4.2}, Spider (code generation)~\cite{mm4.3}, PubMedQA (medical question answering)~\cite{mm4.4}, and SQuAD (reading comprehension)~\cite{mm4.5}.

\highlight{Models.} We choose four representative LLMs for validation. Two of them are specifically designed for on-device deployment: Qwen2-0.5B-Instruct~\cite{mm4.19}, Gemma2-2B-it~\cite{mm4.20}. The other two are larger models: ChatGLM3-6B-32k~\cite{mm4.9} and LLaMA3-8B-Instruct~\cite{mm4.8}.

\highlight{Metric.} For all tasks, we use accuracy as the metric. Specifically, for GSM8k, a prediction is considered correct when the final answer matches the actual result. For PubMedQA, a 3-class classification task is deemed correct if it matches the true label. For Spider, we follow the prior work~\cite{www4.4} and assess whether the generated query matches the reference query. For SQuAD, we align the answer with the reference, as in previous work~\cite{www4.15}.
To evaluate execution overhead, we use Floating Point Operations (FLOPs) as the metric, following the prior research~\cite{mm4.15,mm4.16}.

\highlight{Implementation Details.} We conduct experiments with the \textit{Huggingface library}~\cite{huggingface_transformers}. For optimization, we use the widely adopted AdamW~\cite{www4.2} optimizer and a linear learning rate scheduler~\cite{www4.1}. Same as previous work~\cite{www4.1}, we report results of the runs that achieve the highest performance, consistent with real-world practices prioritizing the optimal model. All experiments are conducted on NVIDIA A800 GPUs with 80GB of VRAM.

\highlight{Baselines.} We compare our method against comprehensive baselines, including ideal bounds and representative defenses, categorized by their protection principles (TPTE, PPTE, PSP):
\tikz[baseline=(char.base)]{\node[shape=circle,draw,inner sep=0.2pt] (char) {1};}
\textbf{Lower bound}: (i) \textit{No-shield}, where the adversary accesses the unprotected model directly.
\tikz[baseline=(char.base)]{\node[shape=circle,draw,inner sep=0.2pt] (char) {2};}
\textbf{Upper bound}: (i) \textit{Black-box}, where only model architecture is visible to the attacker, offering the strongest protection.
\tikz[baseline=(char.base)]{\node[shape=circle,draw,inner sep=0.2pt] (char) {3};}
\textbf{TPTE} (task-only protection): (i) NPLO~\cite{mm0.0}.
\tikz[baseline=(char.base)]{\node[shape=circle,draw,inner sep=0.2pt] (char) {4};}
\textbf{PPTE} (partial model protection):
(i) DarkneTZ~\cite{mm4.12};
(ii) SOTER~\cite{mm4.14};
(iii) Serdab~\cite{mm4.17};
(iv) Our baseline, DTE, runs the latter transformer layers within TEE.
\tikz[baseline=(char.base)]{\node[shape=circle,draw,inner sep=0.2pt] (char) {5};}
\textbf{PSP} (parameter shuffling):
(i) ShadowNet~\cite{mm4.13};
(ii) TransLinkGuard (TLG)~\cite{aaai0.0}. 
To adapt these methods to transformer models, we rigorously configure each solution based on its papers.
Specifically, for SOTER, TEE randomly shields 20\% layers. 
For Serdab, the TEE shields the first transformer layer.
ShadowNet obfuscates all the linear transformation layers.
For DarkneTZ, the last transformer layer is put into TEE. 

\highlight{Model Stealing Attack.}
\label{Model Stealing Attack}
As discussed in prior work~\cite{mm0.0}, we identify finetuning attacks as a major threat to edge-protection solutions. The finetuning attack, as detailed in Appendix \ref{Detailed Model Stealing Attack}, is well-defined in prior studies~\cite{mm0.0,aaai0.0}, and involves two main steps. First, the attacker builds a surrogate model using available parameters of the target model from unsecured environments. Second, they train this surrogate model with accessible datasets to perform the target task. 
Notably, due to the generalization ability of LLMs, attackers can steal the model for any target task, where they may possess sufficient training datasets to achieve stealing. Therefore, we assume the attacker has the entire training dataset (100\%), a more stringent condition than the 1\% dataset employed in previous research.

\subsection{Security Evaluation}

\highlight{Security against Unauthorized Usage.} 
\label{Security Guarantee}
In this subsection, we assess CoreGuard's security against unauthorized usage. We first evaluate its ability to prevent direct unauthorized inference. 
Then, we assess its security against MS attacks. Table \ref{experiments_4.4.1} reports the results: in all cases, the unauthorized inference (``Unau'') accuracy is 0\%, demonstrating CoreGuard's strong resistance to unauthorized inference. In terms of MS attacks, the security of CoreGuard is comparable to the \textit{upper bound}, indicating that attackers cannot misuse the foundational capabilities of the protected model for downstream tasks. Specifically, CoreGuard's relative accuracy is $1.17\times$ compared to the black-box baseline, benefiting from the effective protection provided by our proposed permutation protocal.
specifically, the relative accuracy of CoreGuard ($1.17\times$) is similar to that of DTE ($1.18\times$), which protects the same parameters directly using TEE. This suggests that permutation offers a similar security to the strongest protection (i.e., direct protection by TEE).
Regarding other methods, the TPTE solution, NPLO ($9.59\times$) offers no defense (no-shield is $9.58\times$), and PPTE solutions, e.g., DarkneTZ ($8.43\times$), only provide weak protection.
In contrast, PSP methods, TLG ($1.07\times$) and ShadowNet ($1.09\times$), also reach the upper bound.

\begin{table}[t]
\caption{Security assessment of CoreGuard in preventing unauthorized direct inference and model stealing attack. For \textit{direct inference}, we report the authorized (``Auth'') and unauthorized (``Unau'') usage accuracies (\%). For \textit{model stealing attacks}, we report the attack accuracies (\%), \textit{lower attack accuracies indicate stronger defense}.
The last row reports the average attack accuracies of each defense relative to the baseline black-box solutions. The column representing CoreGuard (``Ours'') is highlighted in \textbf{bold}.
}
\renewcommand{\arraystretch}{0.9}
\centering
\resizebox{1\columnwidth}{!}{
\setlength{\tabcolsep}{2pt}
\begin{tabular}{lcccccccccccccc}
\toprule
& & \multicolumn{2}{c}{\textbf{Direct Inference}} & & \multicolumn{9}{c}{\textbf{Model Stealing Attack $\downarrow$}} \\ 
\cline{3-4} \cline{6-15} \noalign{\smallskip} & & Unau $\downarrow$ & Auth & & No-shield & NPLO & Serdab & DarkneTZ & SOTER& TLG & ShadowNet &  DTE & \textbf{Ours} & Black-box \\
\cline{1-15} \noalign{\smallskip}
\multirow{4}{*}{Qwen2-0.5B} & GSM8k & 0.00\scriptsize\textcolor{gray!150}{$\pm$0.00} & 15.51\scriptsize\textcolor{gray!150}{$\pm$1.02} & & 21.53\scriptsize\textcolor{gray!150}{$\pm$1.43} & 20.92\scriptsize\textcolor{gray!150}{$\pm$1.21} & 14.96\scriptsize\textcolor{gray!150}{$\pm$0.88} & 16.81\scriptsize\textcolor{gray!150}{$\pm$1.07} & 12.50\scriptsize\textcolor{gray!150}{$\pm$0.91} & 1.43\scriptsize\textcolor{gray!150}{$\pm$0.04} & 1.34\scriptsize\textcolor{gray!150}{$\pm$0.04} & 2.36\scriptsize\textcolor{gray!150}{$\pm$0.06} & \textbf{2.41}\scriptsize\textcolor{gray!150}{$\pm$0.07} & 1.29\scriptsize\textcolor{gray!150}{$\pm$0.03} \\
& Spider & 0.00\scriptsize\textcolor{gray!150}{$\pm$0.00} & 5.56\scriptsize\textcolor{gray!150}{$\pm$0.62} &  & 28.48\scriptsize\textcolor{gray!150}{$\pm$1.54} & 30.28\scriptsize\textcolor{gray!150}{$\pm$1.73} & 23.90\scriptsize\textcolor{gray!150}{$\pm$1.21} & 26.01\scriptsize\textcolor{gray!150}{$\pm$1.47} & 21.52\scriptsize\textcolor{gray!150}{$\pm$1.03} & 3.31\scriptsize\textcolor{gray!150}{$\pm$0.10} & 3.67\scriptsize\textcolor{gray!150}{$\pm$0.11} & 3.92\scriptsize\textcolor{gray!150}{$\pm$0.12} & \textbf{3.79}\scriptsize\textcolor{gray!150}{$\pm$0.11} & 3.81\scriptsize\textcolor{gray!150}{$\pm$0.11} \\
& PubMedQA & 0.00\scriptsize\textcolor{gray!150}{$\pm$0.00} & 15.50\scriptsize\textcolor{gray!150}{$\pm$1.24} & & 58.00\scriptsize\textcolor{gray!150}{$\pm$2.56} & 56.50\scriptsize\textcolor{gray!150}{$\pm$2.47} & 49.00\scriptsize\textcolor{gray!150}{$\pm$2.02} & 51.50\scriptsize\textcolor{gray!150}{$\pm$2.19} & 47.00\scriptsize\textcolor{gray!150}{$\pm$1.94} & 3.50\scriptsize\textcolor{gray!150}{$\pm$0.14} & 4.50\scriptsize\textcolor{gray!150}{$\pm$0.18} & 5.50\scriptsize\textcolor{gray!150}{$\pm$0.22} & \textbf{6.00}\scriptsize\textcolor{gray!150}{$\pm$0.24} & 5.00\scriptsize\textcolor{gray!150}{$\pm$0.20} \\
& SQuAD & 0.00\scriptsize\textcolor{gray!150}{$\pm$0.00} & 16.50\scriptsize\textcolor{gray!150}{$\pm$1.28} & & 30.54\scriptsize\textcolor{gray!150}{$\pm$1.75} & 32.33\scriptsize\textcolor{gray!150}{$\pm$1.89} & 28.42\scriptsize\textcolor{gray!150}{$\pm$1.53} & 29.89\scriptsize\textcolor{gray!150}{$\pm$1.64} & 26.34\scriptsize\textcolor{gray!150}{$\pm$1.47} & 6.81\scriptsize\textcolor{gray!150}{$\pm$0.27} & 5.93\scriptsize\textcolor{gray!150}{$\pm$0.24} & 4.42\scriptsize\textcolor{gray!150}{$\pm$0.18} & \textbf{7.35}\scriptsize\textcolor{gray!150}{$\pm$0.29} & 5.66\scriptsize\textcolor{gray!150}{$\pm$0.23} \\
\cline{1-15} \noalign{\smallskip}
\multirow{4}{*}{Gemma2-2B} & GSM8k & 0.00\scriptsize\textcolor{gray!150}{$\pm$0.00} & 30.10\scriptsize\textcolor{gray!150}{$\pm$2.05} & & 40.94\scriptsize\textcolor{gray!150}{$\pm$2.57} & 40.50\scriptsize\textcolor{gray!150}{$\pm$2.41} & 35.18\scriptsize\textcolor{gray!150}{$\pm$1.96} & 37.07\scriptsize\textcolor{gray!150}{$\pm$2.10} & 32.67\scriptsize\textcolor{gray!150}{$\pm$1.72} & 4.58\scriptsize\textcolor{gray!150}{$\pm$0.18} & 10.81\scriptsize\textcolor{gray!150}{$\pm$0.43} & 4.56\scriptsize\textcolor{gray!150}{$\pm$0.18} & \textbf{3.91}\scriptsize\textcolor{gray!150}{$\pm$0.16} & 1.74\scriptsize\textcolor{gray!150}{$\pm$0.07} \\
& Spider & 0.00\scriptsize\textcolor{gray!150}{$\pm$0.00} & 3.52\scriptsize\textcolor{gray!150}{$\pm$0.38} & & 39.15\scriptsize\textcolor{gray!150}{$\pm$1.71} & 38.83\scriptsize\textcolor{gray!150}{$\pm$1.65} & 24.80\scriptsize\textcolor{gray!150}{$\pm$1.08} & 23.29\scriptsize\textcolor{gray!150}{$\pm$0.98} & 12.81\scriptsize\textcolor{gray!150}{$\pm$0.63} & 0.00\scriptsize\textcolor{gray!150}{$\pm$0.00} & 0.00\scriptsize\textcolor{gray!150}{$\pm$0.00} & 0.00\scriptsize\textcolor{gray!150}{$\pm$0.00} & \textbf{0.00}\scriptsize\textcolor{gray!150}{$\pm$0.00} & 0.00\scriptsize\textcolor{gray!150}{$\pm$0.00} \\
& PubMedQA & 0.00\scriptsize\textcolor{gray!150}{$\pm$0.00} & 10.50\scriptsize\textcolor{gray!150}{$\pm$0.83} & & 69.50\scriptsize\textcolor{gray!150}{$\pm$3.21} & 69.00\scriptsize\textcolor{gray!150}{$\pm$3.12} & 55.50\scriptsize\textcolor{gray!150}{$\pm$2.41} & 60.00\scriptsize\textcolor{gray!150}{$\pm$2.63} & 55.50\scriptsize\textcolor{gray!150}{$\pm$2.32} & 10.50\scriptsize\textcolor{gray!150}{$\pm$0.42} & 7.00\scriptsize\textcolor{gray!150}{$\pm$0.28} & 9.50\scriptsize\textcolor{gray!150}{$\pm$0.38} & \textbf{12.00}\scriptsize\textcolor{gray!150}{$\pm$0.48} & 6.50\scriptsize\textcolor{gray!150}{$\pm$0.26} \\
& SQuAD & 0.00\scriptsize\textcolor{gray!150}{$\pm$0.00} & 43.21\scriptsize\textcolor{gray!150}{$\pm$2.76} & & 63.96\scriptsize\textcolor{gray!150}{$\pm$3.08} & 63.94\scriptsize\textcolor{gray!150}{$\pm$3.07} & 60.82\scriptsize\textcolor{gray!150}{$\pm$2.81} & 61.02\scriptsize\textcolor{gray!150}{$\pm$2.84} & 57.87\scriptsize\textcolor{gray!150}{$\pm$2.63} & 7.91\scriptsize\textcolor{gray!150}{$\pm$0.32} & 6.71\scriptsize\textcolor{gray!150}{$\pm$0.27} & 7.51\scriptsize\textcolor{gray!150}{$\pm$0.30} & \textbf{7.81}\scriptsize\textcolor{gray!150}{$\pm$0.31} & 8.81\scriptsize\textcolor{gray!150}{$\pm$0.35} \\
\cline{1-15} \noalign{\smallskip}
\multirow{4}{*}{ChatGLM3-6B} & GSM8k & 0.00\scriptsize\textcolor{gray!150}{$\pm$0.00} & 37.13\scriptsize\textcolor{gray!150}{$\pm$2.33} & & 55.95\scriptsize\textcolor{gray!150}{$\pm$2.87} & 55.07\scriptsize\textcolor{gray!150}{$\pm$2.74} & 53.67\scriptsize\textcolor{gray!150}{$\pm$2.58} & 54.91\scriptsize\textcolor{gray!150}{$\pm$2.74} & 54.55\scriptsize\textcolor{gray!150}{$\pm$2.71} & 2.84\scriptsize\textcolor{gray!150}{$\pm$0.11} & 0.43\scriptsize\textcolor{gray!150}{$\pm$0.02} & 0.93\scriptsize\textcolor{gray!150}{$\pm$0.04} & \textbf{1.04}\scriptsize\textcolor{gray!150}{$\pm$0.04} & 0.23\scriptsize\textcolor{gray!150}{$\pm$0.01} \\
& Spider & 0.00\scriptsize\textcolor{gray!150}{$\pm$0.00} & 5.15\scriptsize\textcolor{gray!150}{$\pm$0.61} & & 35.81\scriptsize\textcolor{gray!150}{$\pm$1.94} & 37.03\scriptsize\textcolor{gray!150}{$\pm$2.03} & 32.25\scriptsize\textcolor{gray!150}{$\pm$1.64} & 33.81\scriptsize\textcolor{gray!150}{$\pm$1.79} & 33.22\scriptsize\textcolor{gray!150}{$\pm$1.75} & 6.19\scriptsize\textcolor{gray!150}{$\pm$0.25} & 8.31\scriptsize\textcolor{gray!150}{$\pm$0.33} & 8.44\scriptsize\textcolor{gray!150}{$\pm$0.34} & \textbf{7.37}\scriptsize\textcolor{gray!150}{$\pm$0.29} & 7.91\scriptsize\textcolor{gray!150}{$\pm$0.32} \\
& PubMedQA & 0.00\scriptsize\textcolor{gray!150}{$\pm$0.00} & 46.00\scriptsize\textcolor{gray!150}{$\pm$3.12} & & 71.00\scriptsize\textcolor{gray!150}{$\pm$3.34} & 70.00\scriptsize\textcolor{gray!150}{$\pm$3.21} & 63.00\scriptsize\textcolor{gray!150}{$\pm$2.89} & 65.50\scriptsize\textcolor{gray!150}{$\pm$3.04} & 60.50\scriptsize\textcolor{gray!150}{$\pm$2.63} & 10.00\scriptsize\textcolor{gray!150}{$\pm$0.40} & 12.00\scriptsize\textcolor{gray!150}{$\pm$0.48} & 12.00\scriptsize\textcolor{gray!150}{$\pm$0.48} & \textbf{12.50}\scriptsize\textcolor{gray!150}{$\pm$0.50} & 12.00\scriptsize\textcolor{gray!150}{$\pm$0.48} \\
& SQuAD & 0.00\scriptsize\textcolor{gray!150}{$\pm$0.00} & 62.11\scriptsize\textcolor{gray!150}{$\pm$3.47} & & 68.13\scriptsize\textcolor{gray!150}{$\pm$3.58} & 68.21\scriptsize\textcolor{gray!150}{$\pm$3.59} & 66.28\scriptsize\textcolor{gray!150}{$\pm$3.46} & 63.91\scriptsize\textcolor{gray!150}{$\pm$3.32} & 62.61\scriptsize\textcolor{gray!150}{$\pm$3.21} & 8.61\scriptsize\textcolor{gray!150}{$\pm$0.34} & 9.56\scriptsize\textcolor{gray!150}{$\pm$0.38} & 9.42\scriptsize\textcolor{gray!150}{$\pm$0.38} & \textbf{8.98}\scriptsize\textcolor{gray!150}{$\pm$0.36} & 9.15\scriptsize\textcolor{gray!150}{$\pm$0.37} \\
\cline{1-15} \noalign{\smallskip}
\multirow{4}{*}{LLaMA3-8B} & GSM8k & 0.00\scriptsize\textcolor{gray!150}{$\pm$0.00} & 33.11\scriptsize\textcolor{gray!150}{$\pm$2.25} & & 53.07\scriptsize\textcolor{gray!150}{$\pm$2.68} & 53.83\scriptsize\textcolor{gray!150}{$\pm$2.71} & 47.79\scriptsize\textcolor{gray!150}{$\pm$2.36} & 51.31\scriptsize\textcolor{gray!150}{$\pm$2.54} & 49.75\scriptsize\textcolor{gray!150}{$\pm$2.43} & 5.61\scriptsize\textcolor{gray!150}{$\pm$0.22} & 4.15\scriptsize\textcolor{gray!150}{$\pm$0.17} & 6.09\scriptsize\textcolor{gray!150}{$\pm$0.24} & \textbf{6.22}\scriptsize\textcolor{gray!150}{$\pm$0.25} & 4.05\scriptsize\textcolor{gray!150}{$\pm$0.16} \\
& Spider & 0.00\scriptsize\textcolor{gray!150}{$\pm$0.00} & 10.67\scriptsize\textcolor{gray!150}{$\pm$1.03} & & 40.04\scriptsize\textcolor{gray!150}{$\pm$1.94} & 41.73\scriptsize\textcolor{gray!150}{$\pm$2.08} & 38.27\scriptsize\textcolor{gray!150}{$\pm$1.89} & 38.14\scriptsize\textcolor{gray!150}{$\pm$1.87} & 36.63\scriptsize\textcolor{gray!150}{$\pm$1.75} & 0.00\scriptsize\textcolor{gray!150}{$\pm$0.00} & 0.57\scriptsize\textcolor{gray!150}{$\pm$0.02} & 1.40\scriptsize\textcolor{gray!150}{$\pm$0.06} & \textbf{1.08}\scriptsize\textcolor{gray!150}{$\pm$0.04} & 0.22\scriptsize\textcolor{gray!150}{$\pm$0.01} \\
& PubMedQA & 0.00\scriptsize\textcolor{gray!150}{$\pm$0.00} & 29.00\scriptsize\textcolor{gray!150}{$\pm$2.12} & & 77.00\scriptsize\textcolor{gray!150}{$\pm$3.85} & 77.00\scriptsize\textcolor{gray!150}{$\pm$3.85} & 72.50\scriptsize\textcolor{gray!150}{$\pm$3.54} & 72.50\scriptsize\textcolor{gray!150}{$\pm$3.54} & 68.00\scriptsize\textcolor{gray!150}{$\pm$3.21} & 9.50\scriptsize\textcolor{gray!150}{$\pm$0.38} & 10.00\scriptsize\textcolor{gray!150}{$\pm$0.40} & 12.50\scriptsize\textcolor{gray!150}{$\pm$0.50} & \textbf{11.00}\scriptsize\textcolor{gray!150}{$\pm$0.44} & 10.50\scriptsize\textcolor{gray!150}{$\pm$0.42} \\
& SQuAD & 0.00\scriptsize\textcolor{gray!150}{$\pm$0.00} & 73.02\scriptsize\textcolor{gray!150}{$\pm$3.94} & & 75.91\scriptsize\textcolor{gray!150}{$\pm$4.02} & 75.20\scriptsize\textcolor{gray!150}{$\pm$3.98} & 67.92\scriptsize\textcolor{gray!150}{$\pm$3.61} & 73.81\scriptsize\textcolor{gray!150}{$\pm$3.87} & 69.12\scriptsize\textcolor{gray!150}{$\pm$3.42} & 11.94\scriptsize\textcolor{gray!150}{$\pm$0.48} & 9.64\scriptsize\textcolor{gray!150}{$\pm$0.39} & 10.48\scriptsize\textcolor{gray!150}{$\pm$0.42} & \textbf{10.01}\scriptsize\textcolor{gray!150}{$\pm$0.40} & 9.71\scriptsize\textcolor{gray!150}{$\pm$0.39} \\
\cline{1-15} \noalign{\smallskip}
\multicolumn{2}{c}{\textbf{Relative Mean Attack Accuracy}}  & - & - & &  $9.58\times$& $9.59\times$ & $8.48\times$ & $8.43\times$ & $8.09\times$ & $1.07\times$ & $1.09\times$ & $1.18\times$ & \textbf{$1.17\times$} & $1.00\times$ \\
\bottomrule
\end{tabular}
}
\vspace{-3mm}
\label{experiments_4.4.1}
\end{table}

\vspace{-8pt}

\subsubsection*{Security under Other Attack Settings.} 
\vspace{-8pt}

\begin{wrapfigure}{r}{0.55\textwidth} 

\centering
    \vspace{-1.5\baselineskip}
    \includegraphics[width=1\linewidth]{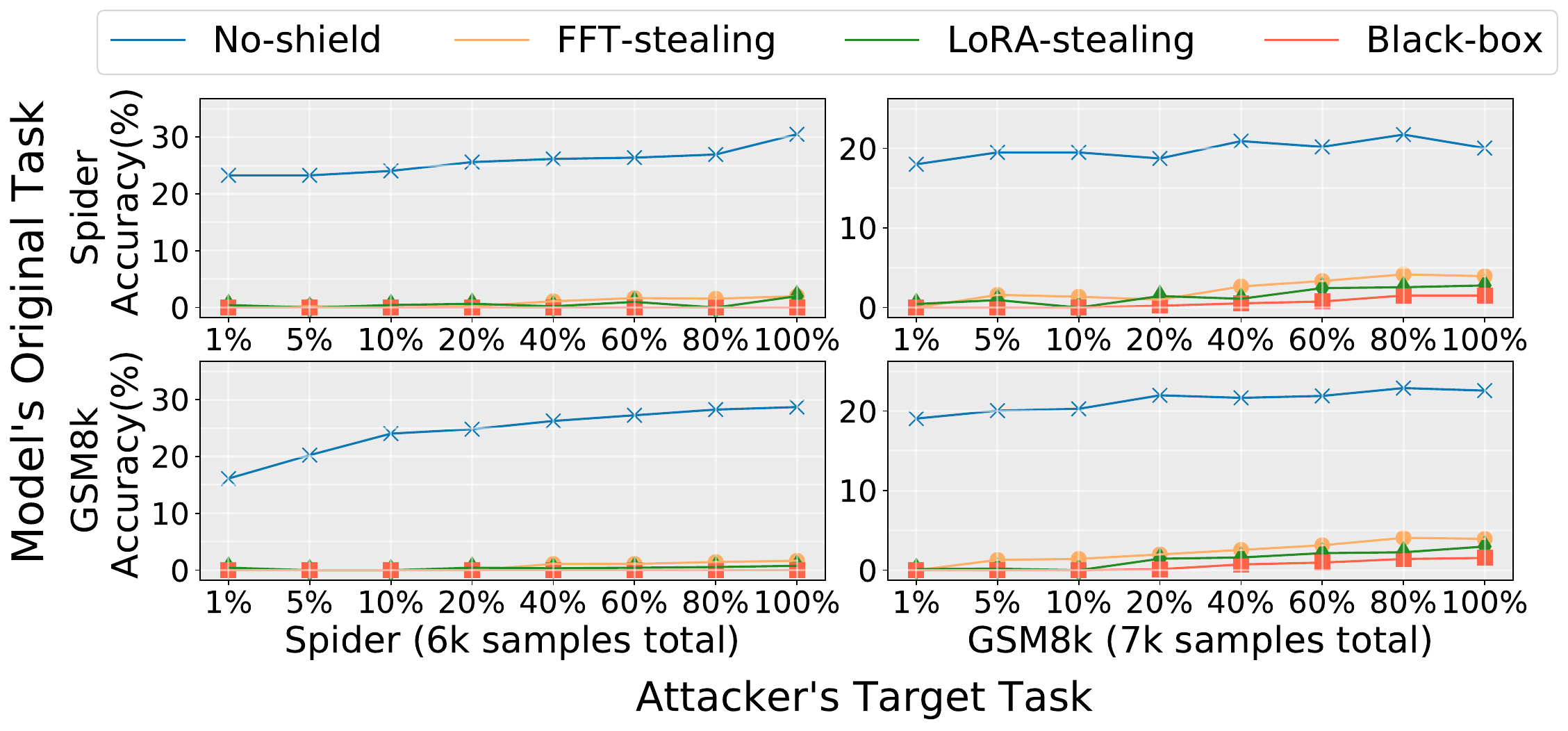}
\vspace{-20pt}
\caption{CoreGuard's Defense Effectiveness Against Model Stealing Across Various Attack Settings.}
\vspace{-8pt}
\label{experiments_4.4.2}
\end{wrapfigure}

In this subsection, we evaluate CoreGuard's security under various attack settings. Specifically, first, prior evaluations focus on FFT for training. To test CoreGuard under different MS training methods, we assess its security against MS using LoRA, a most commonly used LLM fine-tuning approach.
Second, previous evaluations take base models as the deployed model, but in real-world scenarios, LLMs may also be task-customized, which may align with or differ from the attacker’s target, which could affect the defense.
Third, while CoreGuard approaches the upper bound with the entire dataset, attackers may sometimes only have a small portion of the data. To verify whether it can still approach the upper bound in such cases, where the upper bound is also lower, we assess the defense with varying training data, ranging from 1\% to 100\%.
As shown in Figure \ref{experiments_4.4.2}, we report the attack accuracies under various settings on Qwen2-0.5B, and CoreGuard consistently ensures the model's security across all these settings.
The attack accuracies stay close to black-box protection and significantly lower than no-shield cases, regardless of whether the task is aligned, the proportion of data the attacker possesses, or the training method used.

\vspace{-1.5mm}
\begin{tcolorbox}[colback=gray!20!white,colframe=black,left=2mm, right=2mm, top=1mm, bottom=1mm, boxsep=0pt]
\footnotesize
\textbf{Answer to RQ1:} CoreGuard effectively prevents edge-deployed LLMs from being off-device misused, blocking model stealing attacks and achieving upper-bound security across diverse attack settings.
\end{tcolorbox}

\vspace{-1.5mm}

\subsection{Execution and Transfer Overhead}
To answer \textbf{RQ2}, we measure both TEE execution and TEE-GPU transfer overheads to assess CoreGuard's efficiency in computation and communication. 
Specifically, we take an example length of 128 as input and report the TEE execution and data transfer overhead of each solution to generate a single token, excluding TPTE, which offers no protection.
All results are reported in Table \ref{Utility Cost}, and CoreGuard demonstrates a clear advantage.
Specifically, compared to PPTE solutions, CoreGuard achieves thousands of times lower TEE execution overheads. Specifically, the CoreGuard's execution overhead is less than 1.17e-03\% in all cases, whereas PPTE incur execution overheads ranging from 2.91\% to 21.52\%.
More importantly, compared to existing PSP solutions (highlighted in \textbf{bold}), CoreGuard's transfer overhead is nearly two orders of magnitude lower due to its communication-friendly design. Specifically, as mentioned in Section \ref{Inference Authorization}, CoreGuard requires only a single authorization, which is optimal, limiting the transfer rounds to 5. 
In this way, CoreGuard cuts the unacceptable overhead (seconds per token) of existing PSP solutions by two orders of magnitude to negligible levels.

\vspace{-1.5mm}
\begin{tcolorbox}[colback=gray!20!white,colframe=black,left=2mm, right=2mm, top=1mm, bottom=1mm, boxsep=0pt]
\footnotesize
\textbf{Answer to RQ2:} 
CoreGuard's computation and communication overheads are significantly lower than other solutions, showcasing a substantial efficiency advantage.
\end{tcolorbox}

\begin{table*}[t]
\caption{The results of additional overhead. For \textit{execution overhead}, we present the original model's FLOPs (``Original''), the additional overhead in TEE ($FLOPs$), and its proportion to the original model's FLOPs ($\%FLOPs$). For \textit{transfer overhead}, we report the transfer volume ($KB$) and the number of transfers ($rounds$) between the TEE and GPU for each method.}
\vspace{-2mm}
\renewcommand{\arraystretch}{0.9}
\setlength{\tabcolsep}{2pt}
\centering
\resizebox{1\columnwidth}{!}{
\begin{tabular}{lcccccccccccccccc}
\toprule
\multirow{2}{*}{Models} & \multicolumn{8}{c}{\textbf{TEE Execution Overhead $FLOPs/$($\%FLOPs$) $\downarrow$}} & & \multicolumn{7}{c}{\textbf{TEE-GPU Transfer Overhead $KB/rounds$ $\downarrow$}} \\ 
\cline{2-9} \cline{11-17} \noalign{\smallskip}   & \multicolumn{1}{c}{Original} & Serdab  & DarkneTZ &  SOTER & \textbf{TLG} & \textbf{ShadowNet}  & DTE   &  \textbf{Ours} & & \multicolumn{1}{c}{Serdab}  & DarkneTZ &  SOTER & \textbf{TLG} &  \textbf{ShadowNet}  & DTE   &  \textbf{Ours}\\ \midrule
Qwen2                & 1.27E+11          & \multicolumn{1}{c}{\begin{tabular}[c]{@{}c@{}}3.82E+09 \\ (3.02\%)\end{tabular}}        & \multicolumn{1}{c}{\begin{tabular}[c]{@{}c@{}}3.82E+09 \\ (3.02\%)\end{tabular}} & \begin{tabular}[c]{@{}c@{}}2.59E+10 \\ (20.48\%)\end{tabular} & \begin{tabular}[c]{@{}c@{}}\textbf{3.54E+07}  \\ (\textbf{2.79e-02\%})\end{tabular}  & \multicolumn{1}{c}{\begin{tabular}[c]{@{}c@{}}\textbf{3.67E+10} \\ (\textbf{28.97\%})\end{tabular}} & \begin{tabular}[c]{@{}c@{}}4.58E+10  \\ (36.23\%)\end{tabular} & \begin{tabular}[c]{@{}c@{}}\textbf{1.47E+06}  \\ (\textbf{1.17e-03\%})\end{tabular}  
& & \multicolumn{1}{c}{\begin{tabular}[c]{@{}c@{}}2.24E+02 \\ 2\end{tabular}}        & \multicolumn{1}{c}{\begin{tabular}[c]{@{}c@{}}1.12E+02 \\ 2\end{tabular}} & \begin{tabular}[c]{@{}c@{}}5.50E+04 \\ 67\end{tabular} & \begin{tabular}[c]{@{}c@{}}\textbf{1.42E+05}  \\ \textbf{115}\end{tabular} & \multicolumn{1}{c}{\begin{tabular}[c]{@{}c@{}}\textbf{2.75E+05} \\ \textbf{336}\end{tabular}}& \begin{tabular}[c]{@{}c@{}}2.24E+02  \\ 2\end{tabular} & \begin{tabular}[c]{@{}c@{}}\textbf{6.18E+03}  \\ \textbf{5}\end{tabular}
\\ \cline{1-17} \noalign{\smallskip}
Gemma2                & 6.70E+11          & \multicolumn{1}{c}{\begin{tabular}[c]{@{}c@{}}1.99E+10 \\ (2.98\%)\end{tabular}}        & \multicolumn{1}{c}{\begin{tabular}[c]{@{}c@{}}1.99E+10 \\ (2.98\%)\end{tabular}} & \begin{tabular}[c]{@{}c@{}}1.44E+11 \\ (21.52\%)\end{tabular}  & \begin{tabular}[c]{@{}c@{}}\textbf{7.67E+07}  \\ (\textbf{1.14e-02\%})\end{tabular} & \multicolumn{1}{c}{\begin{tabular}[c]{@{}c@{}}\textbf{1.89E+11} \\ (\textbf{28.23\%})\end{tabular}} & \begin{tabular}[c]{@{}c@{}}2.59E+11  \\ (38.72\%)\end{tabular} & \begin{tabular}[c]{@{}c@{}}\textbf{2.95E+06}  \\ (\textbf{4.41e-04\%})\end{tabular} 
& & \multicolumn{1}{c}{\begin{tabular}[c]{@{}c@{}}5.76E+02 \\ 2\end{tabular}}        & \multicolumn{1}{c}{\begin{tabular}[c]{@{}c@{}}2.88E+02 \\ 2\end{tabular}} & \begin{tabular}[c]{@{}c@{}}1.30E+05 \\ 72\end{tabular} & \begin{tabular}[c]{@{}c@{}}\textbf{3.95E+05}  \\ \textbf{125}\end{tabular} & \multicolumn{1}{c}{\begin{tabular}[c]{@{}c@{}}\textbf{6.49E+05} \\ \textbf{364}\end{tabular}} & \begin{tabular}[c]{@{}c@{}}5.76E+02  \\ 2\end{tabular} & \begin{tabular}[c]{@{}c@{}}\textbf{1.58E+04}  \\ \textbf{5}\end{tabular}
\\ \cline{1-17} \noalign{\smallskip}
ChatGLM3                   & 1.53E+12           & \multicolumn{1}{c}{\begin{tabular}[c]{@{}c@{}}5.22E+10 \\ (3.41\%)\end{tabular}}       & \multicolumn{1}{c}{\begin{tabular}[c]{@{}c@{}}5.22E+10 \\ (3.41\%)\end{tabular}} & \begin{tabular}[c]{@{}c@{}}3.01E+11 \\ (19.65\%)\end{tabular} & \begin{tabular}[c]{@{}c@{}}\textbf{2.26E+08}  \\ (\textbf{1.48e-02\%})\end{tabular} & \multicolumn{1}{c}{\begin{tabular}[c]{@{}c@{}}\textbf{4.86E+11} \\ (\textbf{31.75\%})\end{tabular}} & \begin{tabular}[c]{@{}c@{}}7.31E+11 \\ (47.77\%)\end{tabular} & \begin{tabular}[c]{@{}c@{}}\textbf{8.06E+06}  \\ (\textbf{5.27e-04\%})\end{tabular} 
& & \multicolumn{1}{c}{\begin{tabular}[c]{@{}c@{}}1.02E+03 \\ 2\end{tabular}}        & \multicolumn{1}{c}{\begin{tabular}[c]{@{}c@{}}5.12E+02 \\ 2\end{tabular}} & \begin{tabular}[c]{@{}c@{}}1.78E+05 \\ 67\end{tabular} & \begin{tabular}[c]{@{}c@{}}\textbf{5.91E+05} \\ \textbf{135}\end{tabular} & \multicolumn{1}{c}{\begin{tabular}[c]{@{}c@{}}\textbf{1.02E+06} \\ \textbf{336}\end{tabular}} & \begin{tabular}[c]{@{}c@{}}1.02E+03  \\ 2\end{tabular} & \begin{tabular}[c]{@{}c@{}}\textbf{2.19E+04} \\ \textbf{5}\end{tabular}
\\ \cline{1-17} \noalign{\smallskip}
LLaMA3               & 1.92E+12           & \multicolumn{1}{c}{\begin{tabular}[c]{@{}c@{}}5.58E+10 \\ (2.91\%)\end{tabular}}       & \multicolumn{1}{c}{\begin{tabular}[c]{@{}c@{}}5.58E+10 \\ (2.91\%)\end{tabular}} & \begin{tabular}[c]{@{}c@{}}3.85E+11 \\ (20.02\%)\end{tabular} & \begin{tabular}[c]{@{}c@{}}\textbf{1.51E+08}  \\ (\textbf{7.86e-03\%})\end{tabular} & \multicolumn{1}{c}{\begin{tabular}[c]{@{}c@{}}\textbf{5.03E+11}  \\ (\textbf{26.15\%})\end{tabular}} & \begin{tabular}[c]{@{}c@{}}8.94E+11  \\ (46.50\%)\end{tabular} & \begin{tabular}[c]{@{}c@{}}\textbf{4.72E+06}  \\ (\textbf{2.46e-04\%})\end{tabular} 
& & \multicolumn{1}{c}{\begin{tabular}[c]{@{}c@{}}1.02E+03 \\ 2\end{tabular}}        & \multicolumn{1}{c}{\begin{tabular}[c]{@{}c@{}}5.12E+02 \\ 2\end{tabular}} & \begin{tabular}[c]{@{}c@{}}2.82E+05 \\ 90\end{tabular} & \begin{tabular}[c]{@{}c@{}}\textbf{6.36E+05} \\ \textbf{155}\end{tabular}& \multicolumn{1}{c}{\begin{tabular}[c]{@{}c@{}}\textbf{1.31E+06} \\ \textbf{448}\end{tabular}} & \begin{tabular}[c]{@{}c@{}}1.02E+03  \\ 2\end{tabular} & \begin{tabular}[c]{@{}c@{}}\textbf{2.05E+04} \\ \textbf{5}\end{tabular}
\\ 
\bottomrule
\end{tabular}
}
\vspace{-2mm}
\vspace{-0.8\baselineskip}
\label{Utility Cost}
\end{table*}

\begin{table}[t]
\caption{
\sethlcolor{lightblue}
The accuracy comparison between the original model ($M_{ori}$) and the CoreGuard-protected model ($M_{loc}$). The result is presented as $M_{ori}$/$M_{loc}$. Cells showing changes in accuracy are highlighted in \textbf{bold}.}
\renewcommand{\arraystretch}{0.9}
\centering
\resizebox{0.6\columnwidth}{!}{
\begin{tabular}{lcccc}
\toprule
 & GSM8k & Spider & PubMedQA & SQuAD \\
\midrule
Qwen2 & \textbf{15.51\%/15.50\%} & 5.56\%/5.56\%  & 15.50\%/15.50\%  &  16.50\%/16.50\% \\
Gemma2 & 30.10\%/30.10\% & 3.51\%/3.51\%  & 10.50\%/10.50\%  &  43.21\%/43.21\% \\
ChatGLM3 & 37.13\%/37.13\% & 5.15\%/5.15\%  & 46.00\%/46.00\%  &  \textbf{62.11\%/62.09\%} \\
LLaMA3 & \textbf{33.11\%/33.13\%} & 10.67\%/10.67\%  & \textbf{29.00\%/28.50\%}  &  \textbf{73.02\%/73.01\%} \\
\bottomrule
\end{tabular}
}
\vspace{-6mm}
\label{Accuracy Loss}
\end{table}

\vspace{-1mm}
\subsection{Accuracy Loss}
\sethlcolor{lightblue}
To answer \textbf{RQ3}, we compare the accuracy between the unprotected model $M_{ori}$ and the CoreGuard-protected model $M_{loc}$.
The result is shown in Table \ref{Accuracy Loss}. As demonstrated, the impact of CoreGuard on accuracy is minimal.
Specifically, in most cases, there is no difference in accuracy between $M_{ori}$ and $M_{loc}$. However, for some specific cases, accuracy slightly fluctuates (highlighted in \textbf{bold}).
For example, with LLaMA3 on PubMedQA, accuracy decreases slightly by 0.5\%. However, interestingly, we observe a 0.02\% improvement on GSM8k.
Therefore, we consider the minor fluctuations caused by precision limitations (e.g., floating-point errors) rather than the defense itself, which is inevitable.

\vspace{-1.5mm}
\begin{tcolorbox}[colback=gray!20!white,colframe=black,left=2mm, right=2mm, top=1mm, bottom=1mm, boxsep=0pt]
\footnotesize
\textbf{Answer to RQ3:} 
While significantly outperforming existing defenses in terms of both security and efficiency, CoreGuard maintains the model's accuracy without compromise. 
\end{tcolorbox}

\vspace{-2mm}
\section{Limitation and Discussion}
\label{limitation}
\highlight{Side Channel Attacks.}
CoreGuard uses TEEs as its security root, making it vulnerable to side-channel attacks~\cite{aaai6.1,www4.6}.
However, various defense methods have emerged in recent years to mitigate the risk of side-channel leaks, and both these software-~\cite{www4.7,www4.8} and hardware-based~\cite{www4.9,www4.10} defenses can be integrated into our approach. For software-based defense, CoreGuard uses TEE only for basic matrix operations (e.g., matrix permutation and addition). For hardware-based defense, CoreGuard does not require modifications to hardware, allowing physical defense measures to be compatible.

\highlight{TEE in GPUs.}
Recent work has explored implementing trusted environments directly within GPUs~\cite{www4.12}.
However, such GPU/NPU TEEs are still in their early stage and mainly target datacenter settings requiring high-end hardware~\cite{mm5.3}, making them impractical for current edge deployment.
Orthogonal to these solutions, CoreGuard instead focuses on broadly available edge devices—such as smartphones and personal computers—where TEEs are typically CPU-based (e.g., ARM TrustZone, Intel SGX).

\highlight{Real-World Environments.} 
This paper focuses on the core framework, without delving into device-specific implementations. However, this limitation does not affect the general applicability of our approach. CoreGuard is designed to be hardware-agnostic in both implementation and evaluation. In practice, the TEE masks features, performs permutation, and returns the authorized features, relying only on basic TEE functions like matrix operations and data storage, which are universally supported. Additionally, our evaluation is platform-agnostic and suitable across different platforms. 

\vspace{2pt}
\highlight{Architecture Compatibility.} CoreGuard is compatible with mainstream transformer architectures, including LLaMA variants and models with Mixture-of-Experts (MoE) design. For MoE-based transformers, CoreGuard protects each expert independently using the same method applied to standard FFNs. The gating mechanism, typically a linear projection, also supports permutation.

\section{Conclusions}
In this paper, we presents CoreGuard, a protection method that uses a permutation strategy to secure edge-deployed models with maximum security.
Importantly, to reduce transfer overhead during authorization, CoreGuard proposes a propagation protocol, thus only a single authorization is required to authorize the entire model, which is optimal.
Experimental results show that CoreGuard delivers superior security and efficiency without accuracy loss.
In conclusion, CoreGuard is an effective solution that provides model owners with the means to safeguard their proprietary LLMs.

\section*{Acknowledgements}
This work was sponsored by CCF-Huawei Populus Grove Fund. This work was also supported by the Key Project of the National Natural Science Foundation of China under Grant no. 62536007, the Zhejiang Province Science Foundation under Grant no. LD24F020002 and the Zhejiang Province's 2025 "Leading Goose + X" Science and Technology Plan under Grant no. 2025C02034.

\bibliographystyle{plain}
\bibliography{Reference}

\newpage
\appendix
\section{Appendix / Detailed Model Stealing Attack}\label{Detailed Model Stealing Attack}

We consider finetuning attacks as a key security challenge of existing protection solutions.
The finetuning attack is a commonly used attack model that is widely discussed in previous papers and is a well-defined and reproducible attack that is specifically designed for edge-deployed models. This attack typically occurs in two steps. First, the attacker creates a surrogate model and fills it with all the available parameters from the non-secure world. Then, the attacker trains this surrogate model with the datasets they have access to, attempting to apply the surrogate model to their target task. 

Specifically, the attack consists of two phases: \textit{foundational capabilities stealing} ($P_1$) and \textit{task-specific adaptation} ($P_2$).
In $P_1$, to exploit the foundational capabilities of the locked model ($M_{loc}$), the attack begins by inferring the architecture of $M_{loc}$ through its exposed parts. Following this, a replica model, $M_{rep}$, is constructed with the same architecture as $M_{loc}$. Finally, the attacker transports $M_{loc}$'s exposed weights to the corresponding parts of $M_{rep}$.
In $P_2$, the attacker attempts to fine-tune $M_{rep}$ for their tasks. To this end, one potential approach is to train $M_{rep}$ with the training dataset the attacker possesses. Specifically, We assume the attacker has access to the entire training dataset, making this scenario more challenging than previous works\cite{mm1.15,aaai1.5,mm0.0}, which assume attackers have access to only a small portion (e.g., 1\%) of the data. In the training, we mainly consider a more comprehensive and effective method, namely full-parameter training (FFT).  However, to ensure comprehensiveness, we also consider other training settings, such as LoRA.

\section{Appendix / Authorization Position}\label{Authorization Position}

\highlight{Security under Different Authorization Position.}
\label{Hyper-parameters Setting}
The position of authorization is a hyper-parameter of CoreGuard, and its selection can influence the overall security. To identify the best authorization position, we examine how different authorization positions impact security. Specifically, we place the authorization before various transformer layers and evaluate their security based on model-stealing attacks. 

\begin{figure}[ht]
    \centering
    \vspace{-3mm}
    \includegraphics[width=0.45\linewidth]{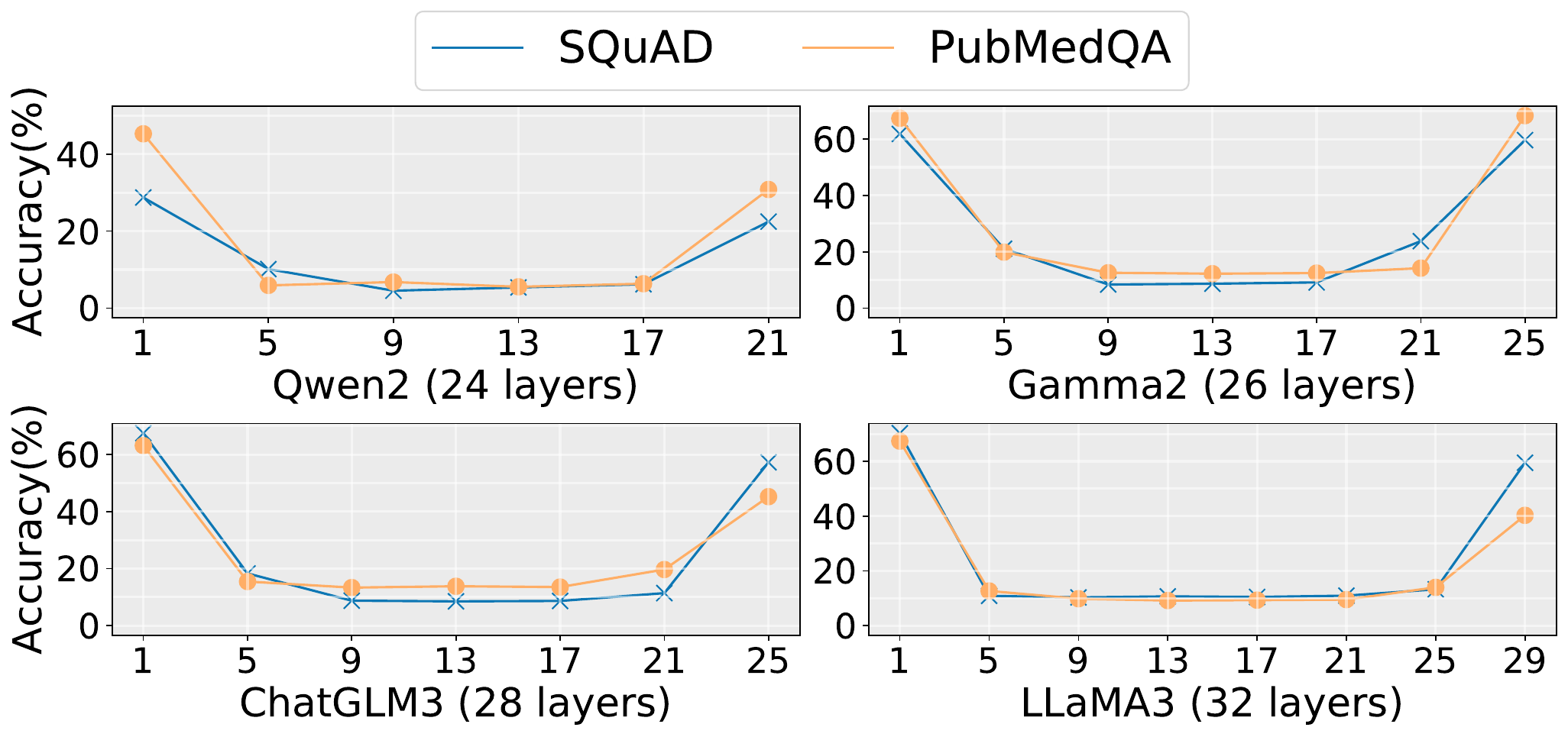}
    \captionsetup{skip=0.4\baselineskip}
    \caption{Impact of authorization position on security. Model-stealing accuracy is reported for different positions, with the total number of transformer layers indicated for each model.}
    \label{authorzation_place}
\vspace{-0.6\baselineskip}
\end{figure}

As shown in Figure \ref{authorzation_place}, the results demonstrate that placing the initial authorization point in the middle of the network is practical. 
Specifically, the model stealing accuracy is higher when the authorization is placed near the beginning or end of the network. 
Conversely, when the authorization is deployed within the central layers of the network, the model-stealing accuracy significantly decreases. This aligns with our expectations: applying authorization at the first layer means CoreGuard permutes nearly all the transformer layers, which means the attacker only needs to recover the functionality of the first transformer layer.
In contrast, placing authorization in the middle leaves at least half of the parameters unaligned, requiring the attacker to recover the functionality of at least half of the network's parameters. Thus, placing the authorization in the central layers is an adequate strategy.

\section{Appendix / Analysis of Neural Network Fitting Attack}\label{NP}

The attacker knows that the input and output are $y,m,m',n',\pi z$. Since Linear A is public, it is meaningless to attack the neural network fitting from $y$ to $m$. In addition, it is meaningless to use $m'$ as input to attack the neural network fitting because $m'$ itself is noisy data. Although $m$ is the same, $m'$ is likely to be different. Therefore, the truly effective input and output of the neural network fitting attack is $m,n',\pi z$.

The attacker's training for neural network fitting attack must be less than the amount of pre-training data, otherwise it will be more trouble than gain. However, the fitted parameters $\hat{W_b},\hat{\pi}$ are definitely noisy. Therefore, the process can be specifically expressed as

\begin{equation}
    m(\hat{\pi} \hat{W_b})+\epsilon = n' \text{ and }  \text{Normalization}(m(\hat{\pi} \hat{W_b}))+\epsilon = \pi z.
\end{equation}

Define $\pi W_b$ as a whole $W$. Then the attacker can crack $\pi$ and $W_b$ as

\begin{equation}
    \hat{\pi} \hat{W_b}+\epsilon=W. 
\end{equation}

In the absence of noise, it is impossible for an attacker to crack it from a mathematical perspective. This is because this is a pathological equation, the number of unknown variables is much greater than the number of equations, and it is a quadratic equation and a nonlinear equation.

From another perspective, the attacker fits infinitely and obtains many $\hat{W_b}$ and $\hat{\pi}$. \textbf{Even if we tell the attacker that the $\hat{W}_b$ is closest to the real $W_b$, it is difficult for the attacker to solve it.} This can be expressed as

\begin{equation}
    \hat{\pi}(W_b-\epsilon')+\epsilon=\hat{\pi}W_b+\tilde{\epsilon}=W,
\end{equation}

where $\epsilon'$ is error between $W_b$ and $\hat{W}_b$. Since floating point numbers have a minimum precision, the equation can be converted to integers by dividing both sides by this minimum precision. Such an integer equation solution problem is a Matrix-Learning With Errors problem.

\paragraph{Hardness of Matrix-Learning With Errors (Matrix-LWE).}
The Matrix-Learning With Errors (Matrix-LWE) problem is a natural generalization of the standard LWE problem. It is defined as follows:
\begin{equation}
    W = \hat{\pi} W_b + \epsilon \mod q.
\end{equation}

If we regard q as the largest positive integer that can be represented by a computer floating point number, then we can directly ignore q. In other words, the attacker's solution to the problem is equivalent to solving the LWE problem.

The goal is to recover the secret \( \hat{\pi} \). Matrix-LWE can be viewed as packing \( \ell \) independent instances of the standard LWE problem:
\begin{equation}
W = \left[ W_b \hat{\pi}_1 + \epsilon_1 \mid W_b \hat{\pi}_2 + \epsilon_2 \mid \cdots \mid W_b \hat{\pi}_\ell + \epsilon_\ell \right] \mod q,
\end{equation}

where each column corresponds to an individual LWE sample with secret \( \hat{\pi}_i \) and noise \( \epsilon_i \).

The hardness of Matrix-LWE follows from the hardness of standard LWE. Regev~\cite{regev2009lattices} proved that solving LWE on average is at least as hard as solving certain worst-case lattice problems such as the Shortest Independent Vector Problem (SIVP\(_\gamma\)) and the Gap Shortest Vector Problem (GapSVP\(_\gamma\)).

Specifically, SIVP\(_\gamma\) asks for a set of \( n \) linearly independent lattice vectors whose maximum length is at most \( \gamma \cdot \lambda_n(\mathcal{L}) \), and GapSVP\(_\gamma\) asks to decide whether the shortest non-zero vector \( \lambda_1(\mathcal{L}) \) in a lattice \( \mathcal{L} \) is smaller than a given threshold \( d \) or larger than \( \gamma d \).

Since these worst-case lattice problems are known to be NP-hard or conjectured to be intractable even for quantum algorithms, Matrix-LWE inherits a strong worst-case to average-case hardness guarantee. In practice, this makes Matrix-LWE a reliable foundation for constructing cryptographic primitives, including post-quantum encryption schemes.

For examply, assume an attacker obtains an approximate estimate \(\tilde W_b\) of the true parameter matrix \(W_b\) satisfying \(\|\tilde W_b - W_b\|_F \le \varepsilon'\), and observes the authorized (locked) output \(W=\pi W_b\). Since \(\pi\) is a permutation matrix in our construction (hence \(\|\pi\|_2=1\)), the attacker can only form
\[
W = \pi \tilde W_b + \varepsilon,\quad\text{with}\quad \varepsilon := \pi (W_b-\tilde W_b),\ \|\varepsilon\|_F \le \varepsilon'.
\]
This matches the canonical Matrix-LWE form \(W=\pi\tilde W_b+\varepsilon\), reducing recovery of \(\pi\) to solving a Matrix-LWE instance.

\section{Appendix / Adaptive Attack}\label{Adaptive Attack}

\highlight{Security against Permutation Matrix Simulation Attack.}
\label{Security against Permutation Matrix Simulation Attack.}
In this subsection, we assess the security of CoreGuard against attackers familiar with the CoreGuard's mechanism and implement attacks accordingly. 
Specifically, the core of authorization involves the permutation matrix (i.e., $\pi$), which the TEE protects.
Therefore, an attacker might first attempt to simulate $\pi$ by initializing a substitute permutation matrix with the same shape and training it based on the TEE's input and output to approximate the true $\pi$. Then, the attacker uses the simulated $\pi$ to mimic the TEE's authorization and fine-tunes the model on their task to complete the attack.
\begin{table}[h]
\vspace{-2mm}
\vspace{-2mm}
\caption{Security evaluation of CoreGuard against permutation matrix simulation attack (\textit{simulation}).}
\renewcommand{\arraystretch}{0.8}
\centering
\resizebox{0.5\columnwidth}{!}{
\begin{tabular}{lcccc}
\toprule
& & \textbf{No-Shield} & \textbf{\textit{Simulation}} & \textbf{Black-box} \\
\cline{1-5} \noalign{\smallskip}
\multirow{2}{*}{Qwen2} & GSM8k& 21.53\%\scriptsize\textcolor{gray!150}{$\pm$1.43\%} &  0.00\%\scriptsize\textcolor{gray!150}{$\pm$0.00\%} & 1.29\%\scriptsize\textcolor{gray!150}{$\pm$0.03\%} \\
 & PubMedQA & 58.00\%\scriptsize\textcolor{gray!150}{$\pm$2.56\%} &  3.50\%\scriptsize\textcolor{gray!150}{$\pm$0.52\%} &  5.00\%\scriptsize\textcolor{gray!150}{$\pm$0.20\%} \\
\cline{1-5} \noalign{\smallskip}
\multirow{2}{*}{Gamma2} & Spider & 39.15\%\scriptsize\textcolor{gray!150}{$\pm$1.71\%} & 0.00\%\scriptsize\textcolor{gray!150}{$\pm$0.00\%}  &  0.00\%\scriptsize\textcolor{gray!150}{$\pm$0.00\%}\\
  & SQuAD & 63.96\%\scriptsize\textcolor{gray!150}{$\pm$3.08\%}  & 0.00\%\scriptsize\textcolor{gray!150}{$\pm$0.00\%} &  8.81\%\scriptsize\textcolor{gray!150}{$\pm$0.35\%} \\
\cline{1-5} \noalign{\smallskip}
\multirow{2}{*}{ChatGLM3} & GSM8k& 55.95\%\scriptsize\textcolor{gray!150}{$\pm$2.87\%} &  0.00\%\scriptsize\textcolor{gray!150}{$\pm$0.00\%} & 0.23\%\scriptsize\textcolor{gray!150}{$\pm$0.01\%} \\
 & PubMedQA & 71.00\%\scriptsize\textcolor{gray!150}{$\pm$3.34\%} &  1.00\%\scriptsize\textcolor{gray!150}{$\pm$0.76\%} &  12.00\%\scriptsize\textcolor{gray!150}{$\pm$0.48\%} \\
\cline{1-5} \noalign{\smallskip}
\multirow{2}{*}{LLaMA3} & Spider & 40.04\%\scriptsize\textcolor{gray!150}{$\pm$1.94\%} & 0.00\%\scriptsize\textcolor{gray!150}{$\pm$0.00\%}  &  0.22\%\scriptsize\textcolor{gray!150}{$\pm$0.01\%} \\
  & SQuAD & 75.91\%\scriptsize\textcolor{gray!150}{$\pm$4.02\%}  & 3.18\%\scriptsize\textcolor{gray!150}{$\pm$0.61\%} &  9.71\%\scriptsize\textcolor{gray!150}{$\pm$0.39\%} \\
\bottomrule
\end{tabular}
}
\vspace{-0.5\baselineskip}
\label{Security against Sophisticated Attackers}
\end{table}

The results are shown in Table \ref{Security against Sophisticated Attackers}; the attack is ineffective, even performing worse than the black-box baseline. The outstanding security is due to the targeted design. Specifically, the non-linear nature of the authorization process, which relies on $\pi$, significantly increases the difficulty of the simulation. Moreover, CoreGuard requires high precision in the authorization process, where even slight simulation errors can compromise model performance.

\highlight{Security against Authorization Simulation Attack.}
\label{Security against Authorization Simulation Attack.}
Considering that precisely fitting $\pi$ is a challenging task, we consider that attackers might attempt to extend their simulation to include adjacent layers or structures, potentially making the attack more feasible. Specifically, since the TEE and the FFN block jointly achieve the authorization, they can be considered as a single unit, which the attacker might attempt to simulate directly. Therefore, the attacker could reconstruct an FFN block structure and train this new FFN block based on the input and output of the original TEE-authorized FFN block, thereby bypassing the TEE's authorization. 
\begin{table}[ht]
\vspace{-2mm}
\vspace{-2mm}
\caption{Security evaluation of CoreGuard against authorization simulation attack (\textit{simulation}).}
\renewcommand{\arraystretch}{0.8}
\centering
\resizebox{0.5\columnwidth}{!}{
\begin{tabular}{lcccc}
\toprule
& & \textbf{No-Shield} & \textbf{\textit{Simulation}} & \textbf{Black-box} \\
\cline{1-5} \noalign{\smallskip}
\multirow{2}{*}{Qwen2} & GSM8k& 21.53\%\scriptsize\textcolor{gray!150}{$\pm$1.43\%} &  2.72\%\scriptsize\textcolor{gray!150}{$\pm$0.61\%} & 1.29\%\scriptsize\textcolor{gray!150}{$\pm$0.03\%} \\
 & PubMedQA & 58.00\%\scriptsize\textcolor{gray!150}{$\pm$2.56\%} &  7.00\%\scriptsize\textcolor{gray!150}{$\pm$1.66\%} &  5.00\%\scriptsize\textcolor{gray!150}{$\pm$0.20\%} \\
 \cline{1-5} \noalign{\smallskip}
\multirow{2}{*}{Gamma2} & Spider & 39.15\%\scriptsize\textcolor{gray!150}{$\pm$1.71\%} & 0.00\%\scriptsize\textcolor{gray!150}{$\pm$0.00\%}  &  0.00\%\scriptsize\textcolor{gray!150}{$\pm$0.00\%} \\
  & SQuAD & 63.96\%\scriptsize\textcolor{gray!150}{$\pm$3.08\%}  & 3.51\%\scriptsize\textcolor{gray!150}{$\pm$0.15\%} &  8.81\%\scriptsize\textcolor{gray!150}{$\pm$0.35\%} \\
\cline{1-5} \noalign{\smallskip}
\multirow{2}{*}{ChatGLM3}   & GSM8k& 55.95\%\scriptsize\textcolor{gray!150}{$\pm$2.87\%} &  0.00\%\scriptsize\textcolor{gray!150}{$\pm$0.00\%} & 0.23\%\scriptsize\textcolor{gray!150}{$\pm$0.01\%} \\
 & PubMedQA & 71.00\%\scriptsize\textcolor{gray!150}{$\pm$3.34\%} &  13.50\%\scriptsize\textcolor{gray!150}{$\pm$0.58\%} &  12.00\%\scriptsize\textcolor{gray!150}{$\pm$0.48\%} \\
 \cline{1-5} \noalign{\smallskip}
\multirow{2}{*}{LLaMA3} & Spider & 40.04\% \scriptsize\textcolor{gray!150}{$\pm$1.94\%}& 0.00\%\scriptsize\textcolor{gray!150}{$\pm$0.00\%}  &  0.22\%\scriptsize\textcolor{gray!150}{$\pm$0.01\%} \\
  & SQuAD & 75.91\%\scriptsize\textcolor{gray!150}{$\pm$4.02\%}  & 6.81\%\scriptsize\textcolor{gray!150}{$\pm$0.72\%} &  9.71\%\scriptsize\textcolor{gray!150}{$\pm$0.39\%} \\
\bottomrule
\end{tabular}
}
\label{authorization simulation attack}
\vspace{-0.4\baselineskip}
\end{table}

As shown in Table \ref{authorization simulation attack}, the attack is ineffective. In all cases, the attack accuracy is similar between the simulation and the black-box baseline but significantly lower than the no-shield baseline. 
The outstanding defense effectiveness is due to the targeted design. Specifically, CoreGuard disrupts the alignment of the parameters before and after the authorization, making it highly challenging for attackers to simply adjust the FFN to recover the compatibility between the two sets of parameters.

\end{document}